# A planar dielectric antenna for directional single-photon emission and near-unity collection efficiency

K. G. Lee[+], X. W. Chen[+], H. Eghlidi, P. Kukura, R. Lettow, A. Renn, V. Sandoghdar, and S. Götzinger

*Laboratory of Physical Chemistry and optETH, ETH Zürich, CH-8093 Zürich, Switzerland*

Abstract: Single emitters have been considered as sources of single photons in various contexts such as cryptography, quantum computation, spectroscopy, and metrology[1,2,3]. The success of these applications will crucially rely on the efficient directional emission of photons into well-defined modes. To accomplish a high efficiency, researchers have investigated microcavities at cryogenic temperatures[4], photonic nanowires[5], and near-field coupling to metallic nano-antennas[6]. However, despite an impressive progress, the existing realizations substantially fall short of unity collection efficiency. Here we report on a theoretical and experimental study of a dielectric planar antenna, which uses a layered structure for tailoring the angular emission of a single oriented molecule. We demonstrate a collection efficiency of 96% using a microscope objective at room temperature and obtain record detection rates of about 50 MHz. Our scheme is wavelength-insensitive and can be readily extended to other solid-state emitters such as color centers[7] and semiconductor quantum dots[8].

One of the most powerful and versatile approaches to the generation of single photons exploits the property that a single quantum mechanical two-level system cannot emit two photons simultaneously since each excitation and emission cycle requires a finite time. Unfortunately, such single-photon sources (SPS) are intrinsically inefficient because their radiation spreads over a $4\pi$ solid angle and cannot be fully captured by conventional optics. Several years ago, a simple avenue for efficient photon collection was proposed by Koyama *et al.* in the context of fluorescence microscopy[9], where emitters were placed at the interface between two media with large refractive index contrast[9,10]. Such a structure can be viewed as a dielectric antenna[11] in which the dipolar radiation of the emitter is funneled into the high-index substrate. The black trace in Fig. 1a shows the angular emission of a dipole sitting close to an interface and oriented perpendicular to it. Despite the strongly modified radiation pattern, one finds that 14% of the light is still lost to the upper half-space, and more importantly, a considerable amount of light is directed to very large angles in the lower substrate, which are not accessible by the collection optics. In this Letter, we remedy these issues by embedding the emitter in a dielectric layer that we engineer on top of the high-index substrate and obtain unprecedented photon collection efficiencies, directionality, and count rates.

To provide an intuitive explanation of our antenna design, let us decompose the radiation of a dipolar emitter into plane waves and consider the propagation of each component[12]. This strategy will also be the basis of the calculations that follow in this work. We start by considering a structure that consists of a lower output substrate with refractive index $n_1$ in contact with an upper medium that contains the emitter and has an index $n_2 < n_1$. To avoid feeding output modes at large angles, we set the distance $h$ between the emitter and the interface to be larger than a characteristic evanescent length[9,10]. This means that the radiation of the emitter into the lower half-space couples to angles $\theta_1 < \sin^{-1}(n_2/n_1)$ in the substrate, but one loses the part that is emitted parallel to the interface and upwards. To get around this problem, we limit the thickness of the upper layer and consider a third medium with refractive index $n_3 < n_2$, as sketched in Fig. 1b. This construction channels the emission of the molecule into quasi-waveguide modes of the middle layer with horizontal wave numbers $k_\rho$ in the range $k_0 n_3 < k_\rho < k_0 n_2$, where $k_0$ is the vacuum wave number. These modes then leak into the substrate at well-defined angles below $\sin^{-1}(n_2/n_1)$ and can be collected with a commercial microscope objective.

As high-index substrate we chose a sapphire cover glass with $n_1$=1.78. A thin film of the polymer polyvinyl alcohol (PVA) with $n_2$=1.50 formed the quasi waveguide, while the top layer was air ($n_3$=1). The red and green curves in Fig. 1a display two examples of out-coupled modes in the $n_1$ medium computed for two choices of PVA thickness (*t*) and emitter-interface distance (*h*). The correspondingly color-coded curves in the inset illustrate the same data in the polar representation. We find that in each case more than 95.5% of the emitted light falls in a cone with a half-angle of 68 degrees, corresponding to a numerical aperture of 1.65. In addition, the contour plot in Fig. 1c shows that the out-coupling efficiency is quite insensitive to variations in *t* and *h*, making the performance of such a device highly tolerant to fabrication imperfections.

As emitters we used single terrylene molecules with fluorescence maximum at 580 nm. To achieve photostability at room temperature, we embedded terrylene in a thin (20 nm) crystalline *p*-terphenyl matrix[13]. This system also has the great advantage that the terrylene molecules are oriented perpendicular to the plane of the *p*-terphenyl film. Our calculations show that an ultrathin film within the middle layer does not influence the theoretical predictions (Fig. 1a), so that we can treat the emitters as dopants directly placed in the second medium, as suggested by Fig. 1b. The schematics of the final device are shown in Fig. 1d. In what follows, we first report on studies of a sample made of two PVA layers that sandwich a *p*-terphenyl film on top of a sapphire cover glass with parameters *t*=350±20 nm and *h*=200±20 nm and also show results for the case of *t*=600±20 nm. The sapphire substrate was index-matched on the other side by an immersion oil suitable for a microscope objective with a numerical aperture of 1.65 (Olympus Apo 100X).

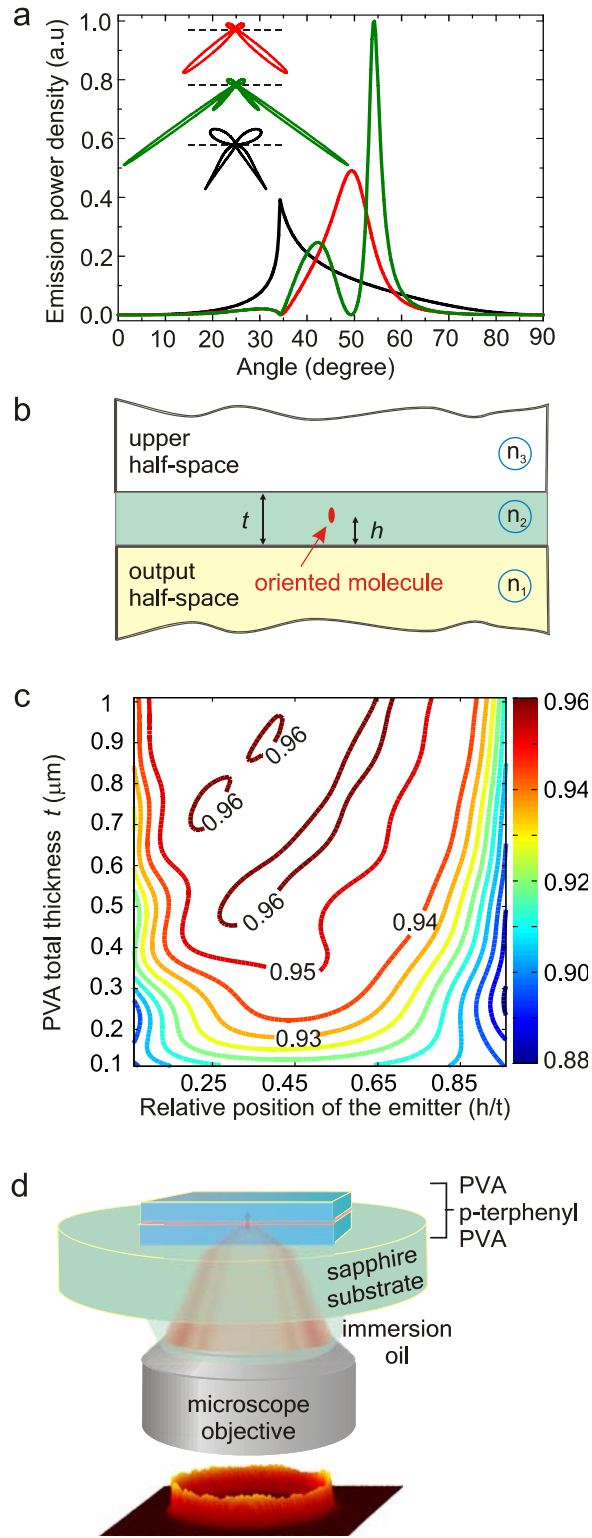

**Figure 1: Emission properties of a vertically-oriented dipole close to a dielectric planar antenna. a** Emitted power density as a function of the collection angle. Black curve: a dipole at a distance of 5nm from the interface of two media with refractive index ratio of 1.78. Red and green: dipoles in a structure as described in (b). The middle layer thickness was $t$= 350 nm and 600 nm, respectively while the emitter was placed at $h$=200 nm in both cases. The insets display the same data in polar coordinates, where the emission into air (upper half space) is enlarged by a factor of 10. **b** Sketch of the layered dielectric antenna. **c**. Contour plot of the fraction of the emitted power that enters the lower substrate within a numerical aperture of 1.65 as a function of $t$ and $h$. **d**. Schematics of the experimental arrangement.

To quantify the emission directivity of our antenna structure, we define the collection efficiency $\eta = S_{co}/S_{em}$ as the ratio of the collected power that enters the microscope objective ($S_{co}$) to the total power emitted by a single molecule ($S_{em}$), both in units of photons per second. The latter is given by the product $N_2 k_{21}$, where $N_2$ is the population of the excited state (2) and $k_{21}$ is the rate of its radiative decay to the ground state (1). As sketched by the energy level scheme in the inset of Fig. 2a, the triplet state (3) also affects the absorption and emission processes. The time trace in the inset of Fig. 2b displays an example of the resulting intermittent on- and off-times. A complete knowledge of all rates and populations requires a thorough study[14], however as we show below, $S_{em}$ can also be evaluated in a less tedious fashion.

The spontaneous emission rate $k_{21}$ is given by the inverse of the rise time ($\tau$) of the autocorrelation function $g^{(2)}$ in the weak excitation limit[15]. At finite excitation powers, $\tau$ is reduced according to $\tau^{-1} = k_{12} + k_{21}$ where $k_{12}$ is the intensity-dependent excitation rate. Figure 2a shows an example of a Hanbury-Brown and Twiss measurement. The red curve is a fit to the experimental data, taking into account the finite time resolution of the avalanche photodiodes and allowing a residual fluorescence background. Figure 2b plots the fluorescence signal as a function of the excitation power. By measuring $g^{(2)}$ and examining $\tau$ for a variety of excitation powers, we determined $k_{21} = 1.26 \times 10^8 \text{s}^{-1}$ and $k_{12}$ for each experiment (see supplementary information). We note that the quantum efficiency of terrylene in *p*-terphenyl is near unity so that the measured excited state decay rate is solely of radiative nature[16].

Published studies [14] and work in our own laboratory reveal that $k_{23}^{-1}$ and $k_{31}^{-1}$ are orders of magnitude longer than $\tau$ (see also on-off times in the inset of Fig. 2b). Consequently, the molecule can be considered as a two-level system during its on-time, allowing us to write $N_2^{on} = k_{21}/(k_{12} + k_{21})$ at steady state ($dN_2/dt = 0$). Hence, the photon emission rate during the on-time can be obtained according to $S_{em}^{on} = N_2^{on} k_{21}$. Now, to compute the long-term total emission rate ($S_{em}^{tot}$), one has to account for the fraction of time the molecule is dark, arriving at $S_{em}^{tot} = N_2^{on} k_{21}(1 - \text{off-time})$. We determined the off-times at different excitation powers directly from recorded time traces (see supplementary information). At the highest excitation power of 7 mW in our experiment (Fig. 2b) we found the off-time to be 5% and $N_2^{on} = 0.82$, so that the emitted power amounted to $0.82 \times (1.26 \times 10^8 \text{ s}^{-1}) \times (0.95) = 9.8 \times 10^7$ photons per second.

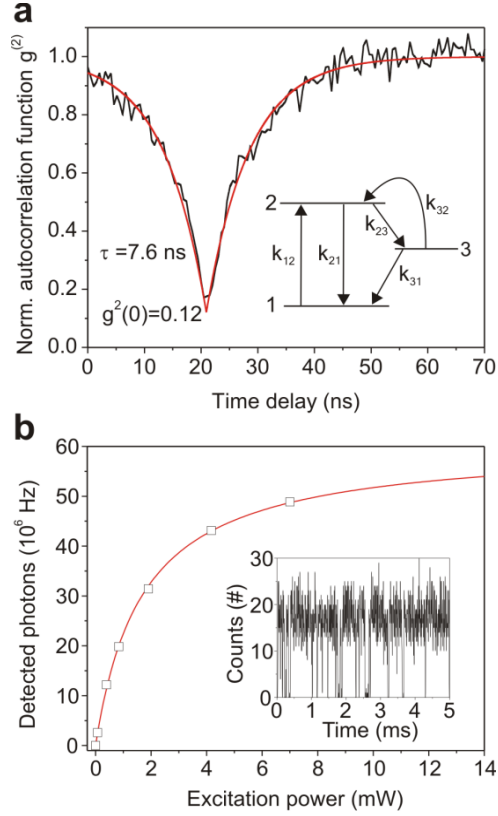

**Figure 2: Photophysics of the single-photon source. a** Second-order photon correlation measurement at an excitation power of 0.07 mW with an integration time of 20 s. At higher excitation power we could measure the second-order correlation function in less than 1 s. The inset shows the simplified level scheme. **b** Fluorescence signal versus the excitation power. The maximum photon detection rate is 4.9*10$^7$ Hz at the highest excitation power of 7 mW. The inset displays a portion of a time-tagged time-resolved experiment with a time bin size of 5 μs.

To deduce $S_{co}$ from the detected count rate ($S_{de}$), we quantified the transmission loss of the optical path and the quantum efficiency of the CCD and found an overall detection efficiency of 51.8% for registering a count per collected photon. Furthermore, we assessed the background level by evaluating the CCD image away from the image spot of the terrylene molecule for each excitation power that was used in the correlation measurements. At the maximum excitation power of 7 mW, we found $S_{co} = 9.4 \times 10^7$, leading to $\eta = 96\% \pm 3\%$. This outcome is in excellent agreement with the theoretically expected value of 96% for a structure with *t*= 350±20 nm and *h*=200±20 nm, sandwiching a *p*-terphenyl film of thickness 20±5 nm. The error accounts for all uncertainties of the setup calibrations and the fitting procedures (see supplementary information).

To examine the angular distribution of the emission pattern and its directionality, we imaged the intensity distribution at the back focal plane of the microscope objective. The inset of Fig. 3a plots the recorded image, while the blue curve in the main figure shows the normalized power distribution as a function of the collection angle (numerical aperture). The red trace in Fig. 3a confirms that the theoretical calculations agree remarkably well with the experimental data. The resulting beam is very close to a radially-polarized doughnut mode, which has been recently discussed in various contexts[17]. It is also possible to convert the observed output modes to other modes such as the fundamental mode of an optical fiber with very high fidelity[18]. In Fig 3b we display the outcome of the same experiment performed for another structure with $t$=600 nm and $h$=200 nm. Again the experiment (blue) reproduces the theoretical predictions (red) very well. We remark in passing that the thicker PVA layer can now support an additional quasi-waveguide mode, giving rise to the second emission lobe.

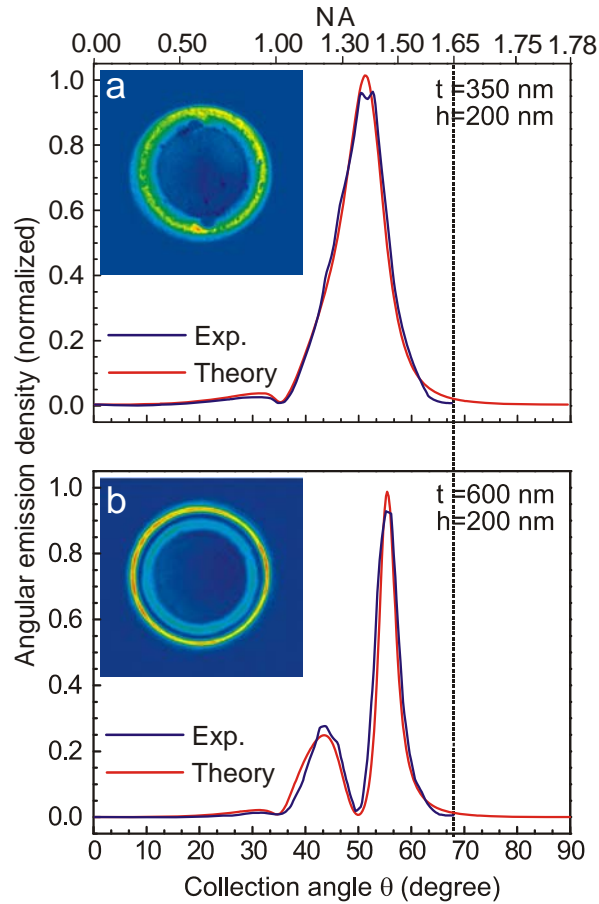

**Figure 3 Angular distribution of the emission. a** Theoretical and measured angular emission intensity for a structure with $t$=350 nm, $h$=200 nm. The inset shows the back focal plane image of the single molecule. **b** Same as in (a) but for a structure with $t$=600 nm, $h$=200 nm. The theoretical curves in both figures account for an angular resolution of 2° imposed by the pixilation of the CCD camera.

Having established a record in collection efficiency, directional emission, and high count rates from a single emitter, we stress the flexibility and potential of our antenna concept for use in solid-state systems. A major advantage is that as opposed to plasmonic structures which suffer from losses and strong dependence on nanoscopic shape and size variations, our design strategy is insensitive to operation wavelength and fabrication imperfections. Thus, the scheme is readily applicable to semiconductor quantum dots, diamond color centers, or solid-state ions alike. Moreover, replacing oil-based microscope objectives by solid-immersion lenses will allow a more compact device both for room-temperature and cryogenic usage. One of the exciting prospects will be the realization of an ultrabright source of Fourier-limited photons[19], which plays a central role for proposals in quantum information science[20]. In particular, we envision using such a SPS for excitation and spectroscopy of a molecule that is embedded in an identical second structure. The modification of the molecular dipolar pattern to a directional emission will allow mode matching between the illumination and the emitter, leading to a highly efficient coupling[21]. Another promise of our bright nonclassical light source is in microscopy and spectroscopy beyond shot noise[1,22] and as a primary intensity standard for metrology[2].

## Acknowledgements

We acknowledge financial support from the Swiss National Foundation (SNF) and the ETH Zurich (QSIT). We thank Mario Agio and Gert Zumofen for helpful discussions and Erkki Ikonen for fruitful exchange regarding the potential of single-photon sources for metrology.